\documentclass[floats,prd,aps,amssymb,twocolumn,nofootinbib]{revtex4}
\usepackage{amssymb}
\usepackage{graphics}
\usepackage{amsmath}
\usepackage{amsfonts}
\usepackage{bm}% bold math
\usepackage{color}
\usepackage[]{latexsym}
\usepackage{amstext}
\usepackage{amssymb}
\usepackage{graphicx}
\usepackage{booktabs}

\usepackage{etoolbox} % for \appto

\newcommand{\be}{\begin{equation}}\newcommand{\ee}{\end{equation}}
\newcommand{\bea}{\begin{eqnarray}}\newcommand{\eea}{\end{eqnarray}}
\newcommand{\brr}{\begin{array}}\newcommand{\err}{\end{array}}
\newcommand{\bit}{\begin{itemize}}\newcommand{\eit}{\end{itemize}}
\newcommand{\ben}{\begin{enumerate}}\newcommand{\een}{\end{enumerate}}

\newcommand{\ba}{\begin{array}}
\newcommand{\ea}{\end{array}}

\def\1{{_{1}}}\def\2{{_{2}}}

\def\noHe0{:\;\!\!\;\!\!:H_e(0):\;\!\!\;\!\!:}
\def\noHm0{:\;\!\!\;\!\!:H_\mu(0):\;\!\!\;\!\!:}

\def\1{{_{1}}}\def\2{{_{2}}}

\def\be{\begin{equation}}
\def\ee{\end{equation}}

\def\bea{\begin{eqnarray}}
\def\eea{\end{eqnarray}}

\def\ve{\varepsilon}

\begin{document}
	
\title{Bekenstein bound and uncertainty relations}

\author{Luca Buoninfante\footnote{buoninfante.l.aa@m.titech.ac.jp}$^{\hspace{0.3mm}1}$, Giuseppe Gaetano Luciano\footnote{gluciano@sa.infn.it}$^{\hspace{0.3mm}2,3}$, Luciano Petruzziello\footnote{lupetruzziello@unisa.it}$^{\hspace{0.3mm}3,4}$ and Fabio Scardigli\footnote{fabio@phys.ntu.edu.tw}$^{\hspace{0.3mm}5,6}$}
 
\affiliation
{\vspace{01mm}$^1$Department of Physics, Tokyo Institute of Technology, Tokyo 152-8551, Japan
\\ 
\vspace{0mm}$^2$Dipartimento di Fisica, Universit\`a di Salerno, Via Giovanni Paolo II, 132 I-84084 Fisciano (SA), Italy
\\ 
\vspace{0mm}
$^3$INFN, Sezione di Napoli, Gruppo collegato di Salerno, Italy
\\
\vspace{0mm}$^4$Dipartimento di Ingegneria, Universit\`a di Salerno, Via Giovanni Paolo II, 132 I-84084 Fisciano (SA), Italy
\\ 
\vspace{0mm}
$^5$Dipartimento di Matematica, Politecnico di Milano, Piazza Leonardo da Vinci 32, 20133, Milano, Italy
\\
\vspace{0mm}
$^6$Institute-Lorentz for Theoretical Physics, Leiden University, P.O. Box 9506, Leiden, The Netherlands}

%\date{\today}

\begin{abstract}
The non--zero value of Planck constant $h$ underlies 
the emergence of several inequalities that must be satisfied 
in the quantum realm, the most prominent one being 
Heisenberg Uncertainty Principle. Among these inequalities, 
Bekenstein bound provides a universal limit on the entropy 
that can be contained in a localized quantum system of given  
size and total energy. In this Letter, we explore how Bekenstein 
bound is affected when 
Heisenberg uncertainty relation is deformed so as to accommodate gravitational effects close to Planck scale (Generalized Uncertainty Principle). By resorting to general thermodynamic arguments, and in regimes where the equipartition theorem still holds,
we derive in this way a ``generalized Bekenstein bound''. Physical implications 
of this result are discussed for both cases of 
positive and negative values of the deformation parameter. 
\end{abstract}

 \vskip -1.0 truecm
\maketitle

\textit{To the cherished memory of Jacob Bekenstein}
%\\
%\\
%%%%%%%%%%%%%%%%%%%%%
\section{Introduction}
%%%%%%%%%%%%%%%%%%%%%
%\\
%\textit{\textbf{Introduction.}}---
In 1981 Jacob Bekenstein proposed a universal upper bound on the entropy $S$ of a localized quantum system~\cite{BB}
\be
S\leq \frac{2\pi\hspace{0.2mm}k_B\hspace{0.2mm}R\hspace{0.2mm}E}{\hbar\hspace{0.2mm}c}\,,
\label{bb0}
\ee 
where $E$ is the total energy of the system and $R=\sqrt{A/4\pi}$
its circumferential radius, with $A$ being the area of the enclosing surface.
Clearly, for $\hbar\rightarrow 0$, 
one obtains $S\leq \infty$, 
which tells us that, classically speaking, 
the entropy of a system is unbounded from above.
The result~\eqref{bb0} was the last offspring of a revolutionary decade 
of investigation, which started with the puzzling proposal 
of Bekenstein himself about the entropy 
of a black hole~\cite{BHE}, then the formulation of black hole 
thermodynamics~\cite{BHTh}, and culminated with the renowned
discovery of Hawking thermal radiation~\cite{BHT}. 

A key assumption in Bekenstein's derivation of the 
bound is that the gravitational self--interaction of the 
system can be neglected. Indeed, Eq.~\eqref{bb0} 
does not contain Newton constant $G_N$, even though 
it was obtained in regimes of strong gravity 
with gedanken experiments involving black holes. 
Remarkably, the inequality is exactly saturated by  Schwarzschild black holes, whose entropy is given by $S=k_B A_H/(2 \ell_p)^2$, where  $A_H$ denotes the horizon area and $\ell_p=\sqrt{\hbar G_N/c^3}$  the Planck length.  
Although many arguments~\cite{BBmany} have been suggested to support the validity of 
Eq.~\eqref{bb0}, also several counterexamples have been brought forward, thereby enriching a lively debate which is still ongoing~\cite{CEX}.
Further years of intuitions and studies have then led to 
the formulation of the well-known Holographic Principle~\cite{GonzalezDiaz:1983yf,tHooft93,Susskind95}, the Covariant~\cite{bousso99} and Causal~\cite{Veneziano2000} Entropy Bounds, and finally to the rigorous quantum field theoretical proof of Bekenstein bound in flat spacetime~\cite{casini}. 
For a general influence of the ideas of Bekenstein on quantum information theory, we remand the reader to Refs.~\cite{BSciAm03,Oppenheim,Wehner}. Connections of various entropy bounds with cosmology~\cite{Veneziano2000,FischlerSusskind98,BanksFischler18,Veneziano99,Bousso16}, perturbative unitarity~\cite{Dvali:2020wqi} and the Pauli principle~\cite{Acquaviva:2020qbc}
have also been addressed with non-trivial results. 

In the last four decades, predictions from
string theory, loop quantum gravity, 
deformed special relativity, non-commutative geometry and 
black hole physics~\cite{Veneziano87,kempf,FS,Adler2,lambiase,AdSa,SC,SC95,Scard18,Luciano19,cas2,qgd,Bossopas} 
have converged on a feasible generalization of Heisenberg Uncertainty Principle (HUP),
which is expected to simultaneously account for quantum
and gravitational effects at Planck scale. In this framework, the standard uncertainty relation for a quantum system should be modified as follows
%usually known as Generalized Uncertainty Principle (GUP), which is expected to take into account also gravity effects at the Planck scale. 
%simultaneously account for quantum and gravitational effects.  In this framework, the standard uncertainty  relation for a quantum system should be modified as follows
\be
\label{gup}
\Delta x\Delta p \geq \frac{\hbar}{2}\left[1 + \beta \left(\frac{\Delta p}{m_p c}\right)^2\right],
\ee
where $\Delta x$ and $\Delta p$ are the position and momentum
uncertainties of the system, respectively, $m_p=\hbar/\ell_p c$ is 
the Planck mass and $\beta$ the so called deformation parameter, 
which is considered to be of order unity in most of quantum gravity models~\cite{SLV}.
The inequality~\eqref{gup} is commonly known as Generalized Uncertainty Principle (GUP).
One of its most important implications is the
significant modification of the behavior of $\Delta x$
as a function of $\Delta p$ in the regime $\Delta p/(m_p c)\simeq1$. 
This results in the prediction 
of a minimum observable length $\Delta x\sim \sqrt{\beta}\hspace{0.2mm}\ell_p$
occurring for $\beta>0$~\cite{FS,Adler2}. 
However, scenarios with $\beta<0$
have been extensively discussed~\cite{JKS,Ong,Buon}, 
along with various remarkable consequences. 
To further substantiate the soundness of
the GUP framework not only at the theoretical level, 
we also mention that several
experiments have been carried out or proposed 
to test the effects predicted by Eq.~\eqref{gup} (see e.g. Refs.~\cite{Experiments}). Of course the GUP should not be intended as a complete theory of Quantum Gravity, fully valid at the Planck scale, neither it claims to be so. A prudent attitude, underlying the most wise literature on this topic, interprets the GUP as an instrument able to describe physics at energies closer to Planck scale, better than what the standard HUP can do. 

Let us remark that Bekenstein bound has been rigorously proved by assuming standard principles of quantum mechanics and quantum field theory in flat spacetime~\cite{casini}. In this Letter, we are interested in understanding how such an entropy bound is affected when HUP is replaced by the GUP in Eq.~\eqref{gup}.
%either some of the fundamental principles are modified and/or the spacetime is curved. Remarkably, by assuming the validity of a GUP, 
%we can somehow contemplate both possibilities. 
To this aim, first we show how Bekenstein inequality~\eqref{bb0} 
can be directly connected to HUP on the basis
of general thermodynamic arguments. 
The present derivation, being elementary and based on first principles, should make it clear why 
Bekenstein bound has such a wide range of validity.
The obtained result is then generalized to the context of GUP, 
leading to a {\it generalized Bekenstein bound} which by construction takes into account also quantum gravitational effects.
Physical implications are finally 
investigated for both positive and negative values of the deformation parameter $\beta$, 
highlighting the different predictions of the two settings. 

%The work is structured as follows: in Sec.~\ref{sec-hup} we  discuss the connection between the HUP and the Bekenstein inequality~\eqref{bb0}.  In Sec.~\ref{sec-gup} the above considerations are extended to the GUP framework.  Conclusions and outlooks are summarized in Sec.~\ref{sec-concl}. 
%Furthermore, from now on
%we shall set $c=1=k_B$ but keep $G_N$ and $\hbar$ explicit.    

%%%%%%%%%%%%%%%%%%%%
\section{Bekenstein Bound and HUP}%\label{sec-hup}
%%%%%%%%%%%%%%%%%%%%

%\textit{\textbf{Bekenstein Bound and HUP.}}--- 
Let us consider an isolated quantum system  
localized inside a finite region of circumferential radius $R$. 
%\textcolor{blue}{From the basics of thermodynamics, it is well-known that
%if the entropy $S$ is a function of the energy $E$ and volume $V$, {\it{i.e.},} $S\equiv S(E,V),$ the temperature satisfies the relation}
From the basics of thermodynamics, 
it follows that if the relation
between the energy $E$, the entropy $S$ and the volume
$V$ of the system is known, then
its temperature $T$ can be easily calculated as follows
%
%\be
%\label{fp}
%T\hspace{0.2mm}{\rm d}S\,=\,{\rm d}E\,,
%\ee
%
\be
\frac{1}{T}=\left(\frac{\partial S}{\partial E}\right)_V\,.
\label{TS}
\ee
By establishing Eq.~\eqref{TS}, we explicitly exclude systems that may possess a negative temperature, otherwise it would result problematic to even introduce the elementary assumptions listed in what follows. Of course, 
Eq.~\eqref{TS} entails also the differentiability of the function $S(E,V)$.
%where we have set ${\rm d}V=0.$ 
%Note that such an assumption encompasses a variety of physical frameworks, since it applies to systems of fixed size, as for example
%a gas of particles enclosed in a rigid container, but also to systems localized in an ideal sphere of radius $R$ at a given time. In fact, 
%Note that, in order to keep our analysis as general as possible, we are not imposing any restriction on the nature of either the system under considerationor its elementary constituents. 
%\footnote{Essentially, we are imaging of packing an energy $E$ in a region of given size $R$. Note that, in general, $R$ does not depend on $E$. It would, if we keep fixed the kind of matter we are packing in, i.e. if we fix a priori the density of such a system. But in principle, we can always think of changing the density of the matter/energy we are packing inside $R$, so that $E$ can increase indefinitely, even maintaining $R$ constant. The crucial question is: can this process be continued at will? Evidently, there is a natural limit: when the gravitational radius associated with the energy $E$ equals or exceed the given size $R$ of the region. At that point, we have created a black hole, and from that point onward, the size $R$ will grow linearly with the total energy $E$ packed inside $R$.}. 

Henceforth, the main working hypotheses underlying our analysis are: 
\begin{enumerate}
	
\item \label{ass-1} We consider a regime where, on average, the energy $\mu$ of each component of the 
system is approximately given by
\be
\label{equipa}
\mu\,\simeq\,\, k_B\, T\,, 
\ee
according to the equipartition theorem.

%\item The system is weakly-interacting, therefore we can assume that the energy $\mu$ satisfies in good approximation the dispersion relation 
%\be
%\label{disp-rel}
%\mu\,\simeq\, \sqrt{p^2c^2+m^2c^4}\,,
%\ee
%where $m$ and $p$ are the mass 
%and momentum of the single constituent, respectively. 

\item The momentum $p$ of each component of our system
satisfies the de Broglie relation 
\be
\label{debr}
p\,=\, \frac{\hbar}{\lambda}\,,
\ee 
where $\lambda$ denotes the corresponding wavelength.   
\end{enumerate}

Note that the second condition only 
makes sense for intrinsically quantum 
particles. From Eq.~\eqref{debr}, it is a simple text-book exercise
to derive Heisenberg relation between the 
momentum and position uncertainties. 
This is a crucial point in the present analysis, 
since Eq.~\eqref{debr} provides the springboard for the extension of the
Bekenstein result to the GUP framework.  

Concerning the first condition, it is well-known that the equipartition theorem is a classical statement. However, it also holds true that, for a large majority of physical systems in regimes close to the classical one, the energy $\mu$ of each component can be approximately described by $\mu \simeq k_B T$ . In other words, Maxwell-Boltzmann statistics is a good approximation of quantum statistics in most of the systems in semiclassical regimes. For example, a gas of bosons at low frequencies or high temperatures is well described by the standard Maxwell-Boltzmann statistics. 

Now, for the above two prescriptions to be valid and from an inspection of the quantum statistics distribution formula, we can infer that the energy $k_B\, T$ should satisfy the condition 
%we can now estimate the energy $k_B\, T$ as
%
\begin{equation}
\label{rel-T-hup}
k_B T \gtrsim\, \frac{\hbar c}{\lambda}\,= pc.
\end{equation}
Given that our system is completely localized inside a volume of radius $R$,  
the inequality $\lambda\lesssim \,2R$ holds true, so that from Eqs.~\eqref{TS} and~\eqref{rel-T-hup} we obtain
\begin{equation}
\label{rel-1/T}
\frac{\partial S}{\partial E}\,=\,\frac{1}{T}\,\lesssim\, \frac{k_B\lambda}{\hbar c}\,\lesssim\, \frac{2 k_B R}{\hbar c}\,,
\end{equation}
where it is understood that the derivative is taken at constant volume.  
%Accordingly, we have
%%
%\begin{equation}
%\label{rel-1/T-R}
%\frac{\partial S}{\partial E}\,\lesssim\, \frac{2 k_B R}{\hbar c}\,.
%\end{equation}
%%

In general, $R$ and $E$ can be regarded as independent variables, therefore
we can easily integrate the above relation with the condition\footnote{For the sake of completeness, we emphasize that the ansatz $S(E=0)=0$ naturally contains the hidden assumption of a unique ground state.} $S(E=0)=0$, obtaining
\begin{equation}
\label{BB-hup}
S\,\lesssim\,\frac{2\alpha k_B\hspace{0.2mm}R\hspace{0.2mm}E}{\hbar\hspace{0.2mm}c}\,,
\end{equation}
%
%where, in the last step, we have
%restored the Boltzmann constant $k_B$ and
%the speed of light $c$ for comparison with Eq.~\eqref{bb0}.
%However, it may be set up
%by requiring that the wavelength of each particle
%of the system is of the order of the radius $R$ of the enclosing volume, 
%with a proportionality constant given by $\alpha$. 
%We shall henceforth include
%it as a ``calibration factor'', just like in the heuristic derivation of the
%Hawking temperature presented in Ref.~\cite{AdSa}. 
where we have inserted a ``calibration factor''
$\alpha$ in order to account for all the approximations
performed so far. Note that this factor cannot be exactly fixed by 
our thermodynamic argument. However, the magnitude of the calibration factor will be obtained in the next Section by means of consistency arguments. Indeed, in analogy with the derivation of the modified Hawking temperature in Ref.~\cite{AdSa}, $\alpha$ can be chosen {\it a posteriori} by requiring that the generalized entropy bound obtained in the GUP framework recovers Bekenstein inequality~(\ref{bb0}) for a vanishing deformation parameter $\beta$ (see below). This occurs for $\alpha=\pi.$

Remarkably, the above 
considerations and the ensuing bound~\eqref{BB-hup} 
%have a wider range of validity, since they 
also encompass the case in which $R$ and $E$ are related via an equation of state.
In fact, for a general and physically plausible radius-energy relation of the form $R=R(E)$, with $R(E)$ being a monotonically non-decreasing function of $E$, one can prove that the inequality~\eqref{BB-hup} is still satisfied (see the Appendix for the proof). 
%this includes systems of constant energy density, $R\sim  E^{1/3},$ and black-hole horizons, $R\sim E.$ 
%We leave a more general study including generic dependences $R\equiv R(E)$ for future.} 

Let us also mention that the Bekenstein bound can be saturated for a system composed by soft quanta, {\it i.e.}, of wavelength $\lambda\sim 2R$. According to the corpuscular models~\cite{Dvali:2011aa,Giusti:2019wdx,Cadoni:2018dnd,Buoninfante:2020tfb}, this can represent the case of a black hole whose constituents are soft gravitons of energy  $\mu\sim \hbar c/\lambda\sim \hbar c/R_s,$ with  $R_s=2G_NM/c^2$ being Schwarzschild radius. 
%In fact, one way to fixrequiring the saturation of (\ref{BB-hup}) for a black hole, which gives $\alpha = \pi$. For such a value of $\alpha$, inequality (\ref{BB-hup}) coincides with inequality  .

Before turning to the calculation of GUP corrections, we stress that 
our result~\eqref{BB-hup}
has been derived by relying on quite general hypotheses. 
Furthermore, we have made no explicit reference to the
particular behavior of the entropy as a function of the energy and/or the number of the elementary constituents. 
Less complete attempts to trace the Bekenstein bound back to HUP can be found in Refs.~\cite{Volovich99,Custodio03}.

It is worth mentioning that the inverse implication, {\it i.e.,} a derivation of HUP from the Bekenstein bound, can also be achieved, as outlined in Ref.~\cite{bousso04}. In a nutshell, let us consider a particle of rest mass $m$ described by a wave-packet
of spatial size $R$, and suppose it is marginally relativistic, namely 
$p\simeq E/c$. For that particle, the inequality (\ref{bb0}) can be recast as
\be
R\,p\, \geq \, \frac{\hbar}{2} \frac{S}{\pi k_B} \,\gtrsim \,\frac{\hbar}{2}\,,
\label{BBHUP}
\ee  
which applies to any system for which $S\gtrsim \mathcal{O}(k_B).$\footnote{For instance, an electron can be in two possible states (spin up and spin down) and therefore its entropy is given by $S=k_B\log 2\sim \mathcal{O}(k_B).$} Of course, the above inequality holds up to a calibration factor which again results equal to $\pi$, but that cannot be determined with this heuristic approach.  

Now, since the direction of motion of our particle is unknown {\it a priori}, we can safely suppose $\Delta p_x \simeq p$, and of course $\Delta x \simeq R$, as for the uncertainty on its position. Therefore Bekenstein inequality (\ref{BBHUP}) can be read as
\be
\Delta x \Delta p_x \gtrsim \frac{\hbar}{2}\,,\label{hup-from-gup}
\ee
which is the standard HUP for the particle in question. 
%Note that a similar derivation could in principle be extended to systems made of different particle species too~\cite{bousso04}. 
Therefore, together with the implication previously shown, the latter argument highlights a full consistency
%establishes a one-to-one correspondence 
between Bekenstein bound and  Heisenberg Uncertainty Principle. %\textcolor{blue}{However, a more rigorous proof, which applies also to non-relativistic particle, is still lacking.}

%%%%%%%%%%%%%%%%%%%%%%%%%%%%%%%%%%%%%%
\section{Generalized Bekenstein bound} %\label{sec-gup}
%%%%%%%%%%%%%%%%%%%%%%%%%%%%%%%%%%%%%%

%\textbf{\textit{Generalized Bekenstein bound.}}--- 
Let us now extend the previous
considerations to the case in which the 
underlying theory is built upon the GUP~\eqref{gup}. 
In particular, we wonder how the
inequality~\eqref{BB-hup} would appear when
taking into account gravity effects
at Planck scale via the GUP. 
Clearly, in order to consistently generalize calculations, 
we need to revise the de Broglie relation in Eq.~\eqref{debr}. 

In the same fashion as HUP 
is in one--to--one correspondence with the de Broglie relation, 
it is reasonable to expect that the GUP is consistent with 
a gravitationally modified de Broglie equation.  This issue has been considered in Ref.~\cite{kempf} and in particular in~\cite{Ahluwalia:2000iw}, where the author obtained a generalized wave-particle duality relation of the form
\begin{equation}
\lambda\,\simeq\, \frac{\hbar}{p}\left[1+\beta \left( \frac{p}{m_p c} \right)^2 \right]. 
\label{gen-de-brog}
\end{equation}
Note that a similar expression is encountered when using the 
GUP in the astrophysical regime, where it gives rise to the so-called ``GUP stars''~\cite{Buoninfante:2020cqz}. 

Equation~\eqref{gen-de-brog} provides the starting point
of our next analysis. By solving it with respect to the momentum $p$, 
we readily obtain
\be
p\,\simeq\,\frac{\hbar\hspace{0.2mm}\lambda}{2\hspace{0.2mm}\beta\hspace{0.2mm}\ell_p^2}\left[1\pm\sqrt{1-4\beta\left(\frac{\ell_p}{\lambda}\right)^2}\right].
\label{mom}
\ee
This reduces to the standard de Broglie relation~\eqref{debr}
in the limit $\beta \ell_p/\lambda\rightarrow 0$ if the negative sign is chosen, 
whereas the positive sign has no evident
physical meaning. Thus, in what follows we only work with the solution corresponding to the minus sign.

We now have all the necessary ingredients to 
derive a generalized Bekenstein bound. Hence, 
by following the same reasoning
as done above, we assume that the 
energy of each quantum constituent 
is given by $\mu\simeq k_B T\gtrsim pc$ and that the system is 
well-localized inside a radius $R$, {\it i.e.}, 
$\lambda\lesssim 2R$.
A comment is here in order. We are still assuming the validity of the equipartition theorem, and considering a regime where $k_BT>pc$. Since we are dealing with the GUP, we are surely closer to Planck energy than what we could reach by describing things just only with the simple HUP. However, we \textit{should not} assume that $pc \sim E_{Planck}$, otherwise this would imply $T>T_{Planck}$, a nonsense. As specified before, the GUP formalism can be trusted for energies enough smaller than $E_{Planck}$, where therefore a regime with $k_BT>pc$ is still imaginable, without running into the oddities of $T \sim T_{Planck}$.    
Thus, the analogue of Eq.~\eqref{rel-1/T} is given by
%%
%\begin{equation}
%\label{rel-T-gup2}
%k_B T\,\gtrsim\,\frac{\hbar\hspace{0.2mm}\lambda c}{2\hspace{0.2mm}\beta\hspace{0.2mm}\ell_p^2}\left[1 - \sqrt{1-4\beta\left(\frac{\ell_p}{\lambda}\right)^2}\right].
%\end{equation}
%%
%As above, we assume
%that the system under consideration is 
%well-localized inside a radius $R$, {\it i.e.}, 
%$\lambda\lesssim 2R$. This yields
%\be
%\label{rew}
%k_B T\,\gtrsim\,\frac{\hbar\hspace{0.2mm}R\,c}{\hspace{0.2mm}\beta\hspace{0.2mm}\ell_p^2}\left[1-\sqrt{1-\beta\frac{\ell_p^2}{R^2}}\right],
%\ee
%
\begin{equation}
\label{rel-1/T-gup}
\frac{\partial S}{\partial E}\,=\,\frac{1}{T}\,\lesssim\,\frac{k_B\,\beta\,\ell_p^2}{\hbar\, R\, c}\left[1-\sqrt{1-\beta\frac{\ell_p^2}{R^2}}
\right]^{-1}\,,
\end{equation}
where we have exploited the fact
that the r.h.s. of Eq.~\eqref{mom}
is a monotonically decreasing function of $\lambda$. Note that Eq.~\eqref{rel-1/T-gup} consistently reduces to Eq.~\eqref{rel-1/T} in the limit $\beta \ell_p^2/R^2 \rightarrow 0$.

%At this stage, let us observe that a crucial r\^ole in deriving our final results is played by the sign of the deformation parameter $\beta$.
In what follows, we discuss separately
the two cases of $\beta>0$ and $\beta<0$.

%%%%%%%%%%%%%%%%%%%%%%%%%%%%%
\subsection{Case $\beta > 0$}
%%%%%%%%%%%%%%%%%%%%%%%%%%%%%

%\textit{\textbf{$\beta > 0$ case.}}--- 
For positive values of the deformation parameter,  
the momentum $p$ in Eq.~\eqref{mom} 
takes real values
%is not complex and unphysical 
only when $\lambda \geq 2\ell_p\sqrt{\beta}$, 
the minimal size allowed by the GUP. 
%In this case, 
%from the condition~\eqref{rew} we are led to
%%
%\begin{equation}
%\label{rel-1/T-gup}
%\frac{\partial S}{\partial E}\,=\,\frac{1}{T}\,\lesssim\,\frac{k_B\,\beta\,\ell_p^2}{\hbar\, R\, c}\left[1-\sqrt{1-\beta\frac{\ell_p^2}{R^2}}
%\right]^{-1}\,,
%\end{equation}
%%
We can now integrate Eq.~\eqref{rel-1/T-gup} under the general 
assumption that $R$ is independent of $E$,
and the usual condition $S(E=0)=0$, thus we obtain
\be
\label{bb-gup-beta>0luciano}
S\,\lesssim\,\frac{\alpha\, k_B\, \beta\,\ell_p^2\,E}{\hbar\, R\, c}\left[1-\sqrt{1-\beta\frac{\ell_p^2}{R^2}}
\right]^{-1}\,,
\ee
that represents the generalized Bekenstein inequality in the case of $\beta>0$. 
Once again, we see that 
the obtained bound is determined up to a factor $\alpha$ 
which can be set by requiring that
Eq.~\eqref{bb0} is recovered in the limit of vanishing $\beta$, and a direct comparison yields 
$\alpha=\pi$. 
Furthermore, as in HUP framework,  Eq.~\eqref{bb-gup-beta>0luciano} 
still holds true for any monotonic non-decreasing radius-energy relation $R=R(E)$. 
%of the form $R\sim E^n,$ with $n$ being a positive real number including zero.

If we now  expand the square root to the next-to-leading order in $\beta \ell_p/R\ll 1$, 
Eq.~\eqref{bb-gup-beta>0luciano} yields
\begin{equation}
\label{bb-gup-beta>0}
S\,\leq\, \frac{2\pi k_B R E}{\hbar c}\left[1-\frac{\beta}{4} \left( \frac{\ell_p}{R} \right)^2 \right]\,,
\end{equation}
where we have inserted the exact numerical factor $\alpha=\pi.$ 
The above relation provides us with the 
effective expression of the generalized Bekenstein bound
in the presently accessible regime, which is far above the Planck scale (we are
assuming $\beta\sim\mathcal{O}(1)$). 

%%%%%%%%%%%%%%%%%%%%%%%%%%%%%
%
\begin{figure}[t]
	\hspace{-2mm}\includegraphics[scale=0.51]{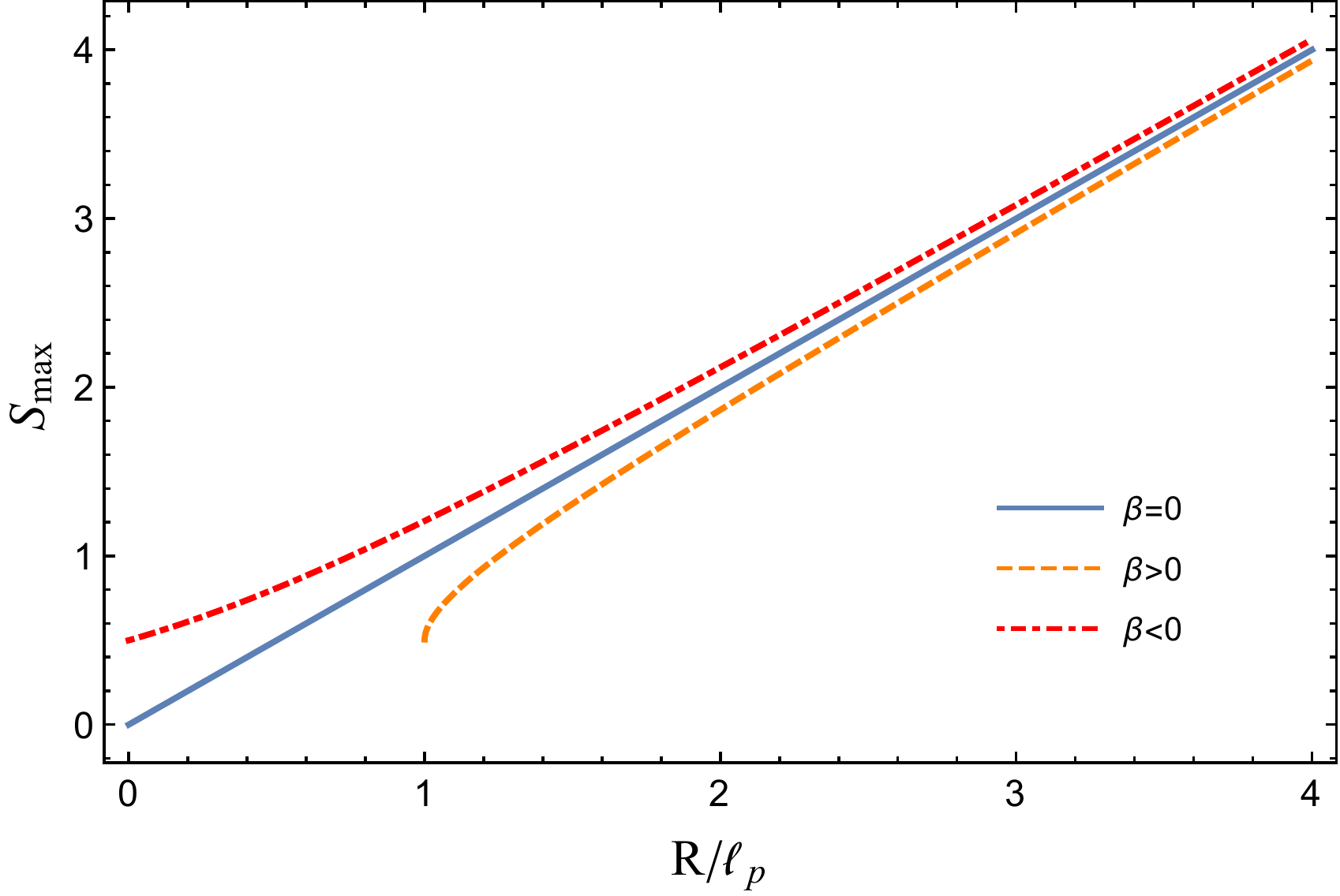}
	\centering
	\protect\caption{Behavior of the upper bound $S_{\rm max}$ as a function of the scaled radius $R/\ell_p$ for $\beta=0$ (blue solid line), $\beta=1$ (orange dashed line) and $\beta=-1$ (red dot-dashed line). We have set $\hbar\hspace{0.2mm} c/(2\pi\hspace{0mm} E\hspace{0.2mm} k_{B}\hspace{0.2mm} \ell_p)=1$. As expected, the discrepancy between the three plots narrows as $R/\ell_p$ increases.}
	%$2\pi E/\hbar=1$ and $\beta \ell_p^2=1$ for simplicity.}
	\label{fig1}
\end{figure}
%%%%%%%%%%%%%%%%%%%%%%%%%%%%%%%%

%%%%%%%%%%%%%%%%%%%%%%%%%%%%%%
%
%\begin{figure*}[t]
	%\includegraphics[width=15cm]{image.png}
	%\centering
	%\protect\caption{Behavior of the upper bound $S_{\rm max}$ as a function of the scaled radius $R/\ell_p$ for $\beta=0$ (black solid line), $\beta=1$ (blue dashed line) and $\beta=-1$ (red dot-dashed line). We have set $\hbar\hspace{0.2mm} c/(2\pi\hspace{0mm} E\hspace{0.2mm} k_{B}\hspace{0.2mm} \ell_p)=1$. As expected, the discrepancy between the three plots narrows as $R/\ell_p$ increases.}
	%%$2\pi E/\hbar=1$ and $\beta \ell_p^2=1$ for simplicity.}
	%\label{fig1}
%\end{figure*}
%%%%%%%%%%%%%%%%%%%%%%%%%%%%%%%%%
The behavior of the generalized Bekenstein bound~\eqref{bb-gup-beta>0luciano}
as a function of the radius $R$
is shown in Fig.~\ref{fig1} (orange dashed line).
Note that the plot 
stops at $R\sim\ell_p$ (we choose $\beta = 1$ for simplicity), consistently with 
the emergence of a minimal length at this scale.
We point out that the GUP correction for $\beta>0$ lowers
the standard Bekenstein limit, thus giving rise
to a more stringent condition on the entropy
which can be stored in a system of given size and total energy.
Consequently, one may then suspect that Schwarzschild black holes
would violate the generalized bound. However, this is not true,
due to the fact that deformations of HUP~\eqref{gup}
affect not only Bekenstein bound, but also the black hole entropy. 
Indeed, if one considers
the GUP-modified expression of the black hole entropy~\cite{AdSa}, it is straightforward
to check that it is still consistent with our bound.

Therefore, from Eqs.~\eqref{bb-gup-beta>0luciano}-\eqref{bb-gup-beta>0}, 
it follows that if a system satisfies the generalized Bekenstein bound, 
it automatically complies with the standard Bekenstein bound too.
In a broader sense, such a result is in line with 
physical intuition. Indeed, it is expected that
the existence of a minimal 
length can reduce the number of microstates 
within a definite
volume, thus decreasing the total amount
of information associated with a system of given size.
In other words, if there is no minimum length,
then one can divide the volume more finely, thus allowing for
higher entropy. 
%We mention that the same result as in Eq.~\eqref{bb-gup-beta>0luciano} has been obtained in Ref.~\cite{Custodio03} by following a different approach.
%Furthermore, a similar reduction of the Bekenstein bound has been found in Ref.~\cite{Kim} in the context of local quantum field theory with a Generalized Uncertainty Principle.
Clearly, for radii $R$ far above the Planck scale, 
GUP effects become increasingly negligible, and in fact
the generalized and standard Bekenstein bounds
tend to coincide.

%%%%%%%%%%%%%%%%%%%%%%%%%
\subsection{Case $\beta<0$}
%%%%%%%%%%%%%%%%%%%%%%%%%

%\textit{\textbf{$\beta < 0$ case.}}--- 
Let us now consider the case of negative 
deformation parameter, $\beta<0$ (which means $\beta=-|\beta|$). 
In this framework there is no
minimal size allowed by the GUP, 
as it can be seen from Eq.~\eqref{gup}. 
Besides this caveat, whose implications
are discussed below, 
calculations and general concepts are
the same as in the previous analysis.

By integrating Eq.~\eqref{rel-1/T-gup} with the generic assumption of $R$ independent from $E,$ we obtain the following upper bound on the entropy
\begin{equation}
\label{bbgup-beta<0}
S\,\lesssim\, \frac{\alpha\, k_B\, |\beta|\,\ell_p^2\,E}{\hbar\, R\, c}
\left[\sqrt{1+|\beta|\frac{\ell_p^2}{R^2}}\,-\,1 \right]^{-1}.
\end{equation}
As before, we set $\alpha=\pi$ by requiring consistency with Eq.~(\ref{bb0}) for $\beta \to 0$. 
Again, inequality (\ref{bbgup-beta<0}) is still true for any relation 
$R=R(E)$ obeying the very plausible property of being monotonic non-decreasing in $E$. 
%\textcolor{blue}{Differently from the previous framework,
%the obtained bound can be extended to the case
%of $R\sim E^n$ provided that the interval $0<n<1$ is ruled out.}
The plot of the new GUP-corrected Bekenstein bound
is shown in Fig.~\ref{fig1} (red dot-dashed line).
For radii $R$ such that $|\beta| \ell_p/R\ll 1$, 
the above expression can be expanded to the next-to-leading order in $\beta$, obtaining 
\be
\label{bb-gup-beta<0}
S\,\leq\, \frac{2\pi k_B R E}{\hbar c}\left[1+\frac{|\beta|}{4}\left( \frac{\ell_p}{R}\right)^2\right]\,,
\ee
which is consistent with Eq.~\eqref{bb-gup-beta>0} with the sign of $\beta$ reversed. 
On the other hand,  the usual Bekenstein bound is recovered for $R \gg \ell_p$, as it should be. 

Now, from a comparison with the $\beta>0$ model, 
we can draw very interesting considerations. 
Indeed, by looking at Eq.~\eqref{bb-gup-beta<0}, 
we immediately notice a striking physical implication: 
because of the positive sign in front of 
the GUP correction, 
the generalized Bekenstein bound with $\beta<0$ 
allows the entropy $S$ of a system 
to exceed the upper limit predicted by Bekenstein.
Of course, this violation is suppressed as
$(\ell_p/R)^2$, so that 
any experimental test appears to be problematic, at least
at present. 
However, we emphasize that such an exotic behavior is
not surprising, if we think that HUP itself can be 
violated for negative values of the deformation parameter. 
Indeed, from Eq.~\eqref{gup}, 
it is clear that, for $\Delta p\sim m_p c$ and $\beta<0$, 
we can have $\Delta x \Delta p \ge 0$, which is typical of a classical regime.
As a matter of fact, the possibility of a quantum-to-classical throwback 
at Planck scale  
has been explored in literature, e.g., by considering $\hbar$ 
as a dynamical field that vanishes in the Planckian limit~\cite{MagSmolin,Hoss}.
Moreover, the scenario in which the universe at Planck energies
appears to be deterministic rather than being
dominated by quantum fluctuations is the vision at the core of `t Hooft's ``deterministic'' quantum mechanics~\cite{thooft,altri,altri2,altri3,altri4}.
In terms of momentum and wavelength, this 
means that the quantum wave-packet of an object with momentum $p\simeq m_p c$ can be maximally localized, {\it i.e.}, $\lambda\simeq 0,$ consistently with the fact that a GUP with $\beta<0$ does not predict any minimal length~\cite{JKS}.

Finally, in connection with the possibility
of accessing arbitrarily short distances in the case of $\beta<0$, 
let us observe that the upper bound in Eq.~\eqref{bbgup-beta<0}  
converges to $\pi k_B\sqrt{|\beta|}E\ell_p/(\hbar c)$
for $R \to 0$. 
This would imply that 
a small - but finite - amount of entropy/information
may be packed in a region of
whatever small size, contrary to intuitive expectations. 
However, such a result 
%(the non-zero value of the entropy
%for systems of vanishing size) 
is most likely just 
a signal that we are trying to extrapolate
our considerations outside their domain of validity.
It is actually evident that, for $R=0$, the energy of the system
cannot but be zero. This means that
$S(R=0)=S(E=0)=0$, according
to the normalization we have adopted. 
Moreover, as shown in Ref.~\cite{JKS}, the GUP with $\beta<0$ seems to be implied by a reticular structure of the spacetime, which would make in any case the limit $R\rightarrow0$ essentially meaningless.
Surely the above aspects
deserve deeper attention and
will be better investigated elsewhere.

%Beyond these speculations, 
%it is worth
%noting that the limit $R\rightarrow 0$
%may be somehow problematic by itself, even for negative
%values of the deformation parameter.
%Indeed, as shown in Ref.~\cite{JKS}, the GUP with $\beta<0$ seems to be 
%connected with a reticular structure of the spacetime, 
%which would make the limit $R\rightarrow0$ essentially meaningless.

%%%%%%%%%%%%%%%%%%%%%%%%%%%%%%%%%%%
\section{Concluding remarks} %\label{sec-concl}
%%%%%%%%%%%%%%%%%%%%%%%%%%%%%%%%%%%

%\textit{\textbf{Concluding remarks.}}--- 
In this Letter we have presented arguments in favour of a full consistency between Heisenberg Uncertainty Principle
and Bekenstein bound on the entropy of a localized system with a given size  and total energy.
Such a result has paved the way for the generalization of the Bekenstein inequality close to the Planck scale, where both quantum and gravitational effects are expected to come into play. 
In particular, we have argued that, if the underlying theory has a Generalized Uncertainty Principle built in, and in regimes where the equipartition theorem still holds, then Bekenstein bound turns out to be non-trivially modified;
corrections 
have been computed in both cases of positive and negative 
values of the deformation parameter, see Eqs.~\eqref{bb-gup-beta>0} and~\eqref{bb-gup-beta<0}, paying great attention to the issue of the
minimal length emerging when $\beta>0$. 
Apart from the well-known Holographic Bound (which is meant to apply to the most general spacetimes of any curvature), to the best of our knowledge this is a {first} concrete attempt towards a derivation of an upper bound on the entropy that takes into account both quantum and gravitational effects close to the Planck scale, thus going beyond the flat-space proof based on standard quantum field theory with canonical commutation relations~\cite{casini}.

Apart from its intrinsic relevance, 
we point out that the obtained result finds applications 
in several other contexts. For instance, it may have
significant implications on
the holographic bound~\cite{GonzalezDiaz:1983yf,tHooft93,Susskind95,bousso99}. 
In fact, by assuming the absence of gravitational instability, or in other words that the size of the system $R$ is larger than the corresponding gravitational radius, 
Eq.~\eqref{bb-gup-beta>0} leads to a generalization of the holographic bound, 
with potential connections
to the world of quantum information theory (see Refs.~\cite{Oppenheim,Wehner}). 
%Finally, we emphasize that our heuristic proof is based on the requirement 
%of fixed radius $R$.
Finally, we expect that the inequality~\eqref{bb-gup-beta>0}, once properly extended to black hole physics, would allow us to establish a link
with the theory of black hole remnants~\cite{ScGrChen,ChenOng}. 
%According to the standard Hawking theory, which is based on the HUP, 
%the evaporation process of a black hole continues until the whole mass is converted into photons, gravitons, and heavier particles. The final temperature reached by this process could be, in principle, infinite. On the other hand, the GUP model with $\beta>0$ 
%predicts that a black hole should stop
%shrinking when its mass is around $M\sim m_p$ (assuming $\beta\sim\mathcal{O}(1)$), at a finite temperature, and therefore
%leaving a thermodynamically inert object called "remnant" behind, 
%which interacts with the environment only gravitationally through its mass--energy. 
Remnants have been thought to be good candidates to model dark matter~\cite{ChenAdler}
%%. Furthermore, having the possibility to store information, they 
and could also play an important r\^ole in 
the resolution of the information loss paradox (see, for instance, Ref.~\cite{ChenOng} and 
therein).
This and further aspects are presently under active investigation and will
be discussed elsewhere.  
%\\
\acknowledgments
L.~B. acknowledges financial support from JSPS and KAKENHI Grant-in-Aid for Scientific Research No. JP19F19324. 
We thank the anonymous Referees for important observations which helped us to improve the quality of the article.

%%%%%%%%%%%%%%%%%%%%%%%%%%%%%%%%%%%
\subsection*{Appendix}  %\label{Appendix}
%%%%%%%%%%%%%%%%%%%%%%%%%%%%%%%%%%%

In this Appendix we show that our derivation of the inequalities (\ref{BB-hup}), (\ref{bb-gup-beta>0luciano}), (\ref{bbgup-beta<0}) holds for any monotonically non-decreasing function $R=R(E)$.
Let $R(\ve)$  and  $g(R)$ be two positive, monotonically non-decreasing functions of $\ve$ (with $0\leq\ve\leq E$) and $R,$ respectively. By introducing the partial derivative $S^\prime(\ve):=\partial S/\partial \ve,$ the inequalities~\eqref{rel-1/T} and~\eqref{rel-1/T-gup} can be written in the following compact form
%(neglecting inessential constants)
\be
S^\prime(\ve) \lesssim g(R(\ve))\,.
\ee
We can now integrate the above inequality with the usual condition $S(\ve=0)=0$ and obtain
\be
S(E) = \int_0^E {\rm d}\ve\,  S^\prime(\ve)  \lesssim \int_0^E {\rm d}\ve\, g(R(\ve)) \leq E\, g(R(E))\,,
\ee
where we used the fact that also $g(R(\ve))$ is a monotonically non-decreasing function of $\ve$ as it is a composition of two monotonically non-decreasing functions.
Therefore, we proved that $S(E) \lesssim E\,g(R(E))$, which resumes the inequalities  
(\ref{BB-hup}), (\ref{bb-gup-beta>0luciano}), (\ref{bbgup-beta<0}).

\end{document}